\documentclass[twocolumn,superscriptaddress,showpacs,aps]{revtex4}

\usepackage{amsmath,amssymb,bm}
\usepackage{mathrsfs,amstext}
\usepackage{hyperref}
\usepackage{graphicx,epstopdf} 
\usepackage{color}

\newcommand{\iChEM}{{\it i}{\rm ChEM}}

\newcommand{\lgter}{\mbox{\tiny $\lessgtr$}}
\newcommand{\tS}{\mbox{\tiny S}}
\newcommand{\B}{\mbox{\tiny B}}

\newcommand{\T}{\mbox{\tiny T}}
\newcommand{\ti}{\Tilde}
\newcommand{\wti}{\widetilde}
\newcommand{\w}{\omega}
\newcommand{\nl}{\nonumber \\}
\newcommand{\nla}{\nl&\quad}

\newcommand{\Sec}[1]{Sec.\;\ref{#1}}
\newcommand{\App}[1]{Appendix\;\ref{#1}}
\newcommand{\be}{\begin{equation}}
\newcommand{\ee}{\end{equation}}
\newcommand{\bsube}{\begin{subequations}}
\newcommand{\esube}{\end{subequations}}
\newcommand{\Eq}[1]{Eq.\,(\ref{#1})}
\newcommand{\Eqs}[1]{Eqs.\,(\ref{#1})}
\newcommand{\Fig}[1]{Fig.\,\ref{#1}}

\newcommand{\dg}{\dagger}
\newcommand{\la}{\langle}
\newcommand{\ra}{\rangle}
\newcommand{\La}{\big\la}
\newcommand{\Ra}{\big\ra}

\newcommand{\cf}{cf.\ }

\usepackage{CJKutf8} 

\begin{document}

\begin{CJK*}{UTF8}{gbsn}

\title{ Memory effect preserved time-local approach to noise spectrum
of transport current
}

\author{Yishan Xu}
\affiliation{School of Physics, Hangzhou Normal University,
Hangzhou, Zhejiang 311121, China}

\author{Jinshuang Jin} \email{jsjin@hznu.edu.cn}
\affiliation{School of Physics, Hangzhou Normal University,
Hangzhou, Zhejiang 311121, China}

\author{Shikuan Wang}
\affiliation{Department of Physics, Hangzhou Dianzi University, Hangzhou 310018, China}

\author{YiJing Yan} \email{yanyj@ustc.edu.cn}
\affiliation{Hefei National Laboratory for Physical Sciences at the Microscale
\& \iChEM,
University of Science and Technology of China, Hefei, Anhui 230026, China}

\date{\today}

\begin{abstract}
Within the second-order non-Markovian
master equation description, we develop an efficient method for
calculating the noise spectrum of transport current through interacting mesoscopic systems.
By introducing proper current-related density operators, we propose a
practical and very efficient time-local approach to compute the noise spectrum,
including the asymmetric spectrum, which contains the full information of
energy emission and absorption.
We obtain an analytical formula of the current
noise spectrum to characterize the nonequilibrium
transport including electron-electron Coulomb interaction
and the memory effect.
We demonstrate the proposed method in transport through interacting-quantum-dots system,
and find good agreement with the exact results under broad range of parameters.

\end{abstract}

\pacs{05.40.Ca, 73.63.-b, 73.23.Hk}
\maketitle

\section{Introduction}
\label{thintro}

 In quantum transport, current correlation function
contains more information than average current 
  \cite{Bla001,Imr02,Bee0337,Naz03}.
Experiments often measure the power spectrum,
the Fourier transformation of correlation function
\cite{Deb03203,Del09208,Bas10166801,Bas12046802,Del18041412}.
In general, nonequilibrium noise spectrum of transport current
neither is asymmetric and nor satisfies
the detail--balance relation
\cite{Eng04136602,Agu001986,
Nak19134403,Ent07193308,Rot09075307,Mao21014104}.
Moreover, mesoscopic systems with discrete energy levels
exhibit strong Coulomb interaction
and the contacting electrodes in general
induce memory effect \cite{Bla001,Imr02,Bee0337,Naz03}.
Theoretical methods that are practical to general mesoscopic systems,
with Coulomb interaction and memory effects on quantum transport,
would be needed.

 As real--time dynamics is concerned,
the quantum master equation approach is the most popular.
Jin--Zheng--Yan established the exact fermionic
hierarchical equation of motion approach \cite{Jin08234703}.
This nonperturbative
theory has been widely used in studying nanostructures
with strong correlations including the Kondo problems
\cite{Zhe09164708,Li12266403,Wan13035129,Che15033009}.
Recently, Yan's group further developed the dissipaton equation of motion (DEOM)
theory \cite{Yan14054105,Jin15234108,Yan16110306,Jin20235144}.
The underlying algebra addresses the hybrid bath dynamics.
The current correlation function can now be evaluated,
even in the Kondo regime \cite{Jin15234108,Mao21014104}.
Note also that
Zhang's group established an exact fermionic master equation,
but only for noninteracting systems \cite{Tu08235311,Jin10083013,Yan14115411}.

In this work, we extend the convention
time-nonlocal master equation (TNL-ME)
to cover an efficient evaluation of
transport current noise spectrum.
The key step is to identify the underlying
current related density operators.
This converts TNL-ME
into a set of three time-local equation-of-motion (TL-EOM) formalism.
The latter has the advantage in such as the initial value problems
and nonequilibrium non-Markovian correlation functions
  \cite{Jin16083038}.
The underlying algebra here is closely related to the DEOM theory
\cite{Yan14054105,Jin15234108,Yan16110306,Jin20235144}.
TL-EOM provides not only real-time dynamics,
 but also analytical formulae for both transport current and noise spectrum.
%

 The remainder of this paper is organized as follows.
In \Sec{thNMK-ME}, we introduce the transport model and
 the energy-dispersed TL-EOM formalism.
In \Sec{thcurr}, combining the TL-EOM and
dissipaton decomposition technique,
we present an efficient method for calculating the current noise spectrum.
The time-dependent current formula is first given in \Sec{thsubcurr}.
 We then derive the current correlation function
 and straightforwardly obtain the analytical formula of noise
 spectrum in \Sec{thsubcurrcf}
 and \Sec{thsubnoise}, respectively.
The detail derivation is given in \App{thappsw}.
We further give discussions and remarks on the resulting noise spectrum formula
in \Sec{thRemarks}.
For illustration, we apply the present method to demonstrate the quantum noise spectra
of the transport through interacting double quantum dots in \Sec{thnum}.
 The numerical results are further compared with
the accurate ones based on DEOM theory.
Finally, we conclude this work with \Sec{thsum}.

\section{Non-Markovian master equation formalisms}

\label{thNMK-ME}

\subsection{Model Hamiltonian}

Consider the electron transport through the central nanostructure system 
 contacted by the two electrode reservoirs (left $\alpha = {\rm L}$ and
 right $\alpha = {\rm R}$).
The total Hamiltonian reads
\begin{align}\label{Htot0}
H_{\T}\!=\!H_{\tS}\!+\!\sum_{\alpha k}\varepsilon_{\alpha k}
 c^{\dg}_{\alpha k}  c_{\alpha k}\!+\!\sum_{\alpha k}\!\left(t_{\alpha u k}a^\dg_u
 c_{\alpha k} \!+\!{\rm H.c.}\right).
\end{align}
The system Hamiltonian $H_{\tS}$ includes electron-electron interaction,
given in terms of local electron creation $ a^{\dg}_{u}$
(annihilation $  a_{u}$) operators of the spin-orbit state $u$.
 The second term describes the two electrodes ($H_{\B}$) modeled by the
 the reservoirs of noninteracting electrons
and $c^{\dg}_{\alpha k}$ ($c_{\alpha k}$) denotes the creation (annihilation) operator
of electron in the $\alpha$-reservoir with momentum $k$ and
energy $ \varepsilon_{\alpha k}$.
The last term is the standard tunneling Hamiltonian between the system and the electrodes
with the tunneling coefficient $t_{\alpha u k}$.
Throughout this work, we adopt the unit of $e=\hbar=1$.

For convenience, we reexpress the tunneling Hamiltonian as
\be\label{Hsb1}
  H'\!=\!\sum_{\alpha u }\left(  a^{+}_{u}    F^-_{\alpha u}
    +   F^+_{\alpha u}   a^{-}_{u} \right)\!=\!\sum_{\alpha u \sigma}  a^{\bar\sigma}_{u}
    \wti F^{\sigma}_{\alpha u},
\ee
where $ F^-_{\alpha u}
=\sum_k t_{\alpha u k} c_{\alpha k}=(  F^+_{\alpha u})^\dg$
and
 $\wti F^{\sigma}_{\alpha u} \equiv \bar\sigma    F^{\sigma}_{\alpha u}$,
with $\sigma =+,-$ ($\bar\sigma$ is the opposite sign to $\sigma$). 
As well-known, the effect of the reservoirs on the transient dynamics of the
central system is characterized
by the bath correlation function,
 \begin{align}\label{ct}
    C^{(\sigma)}_{\alpha uv} (t )
  & \!=\! \la F^\sigma_{\alpha u} (t) F^{\bar\sigma}_{\alpha v} (0)  \ra_{\B}
  \!= \!\int_{-\infty}^{\infty}\!\!\frac{{\mathrm d} E}{2\pi}\,e^{\sigma i E t}
   \Gamma^{\sigma}_{\alpha uv}(E),
  \end{align}
 where $\la \cdots\ra_{\B}$ stands for the statistical average
over the bath (electron reservoirs) in thermal equilibrium.
  The second identity in \Eq{ct} arises from the
    the bath correlation function related to
   the hybridization spectral density
   $J_{\alpha u v}(E)
\equiv2\pi\sum_k t_{\alpha u k}t^\ast_{\alpha v k}\delta(E-\varepsilon_{\alpha k})
=J^\ast_{\alpha vu}(E)$.
Here, we introduced
\be\label{cw-real}
\Gamma^{\sigma}_{\alpha uv}(E)\equiv n^{\sigma}_\alpha(E) J^{\sigma}_{\alpha uv}(E),
\ee
with $J^+_{\alpha vu}(E) = J^-_{\alpha uv}(E) = J_{\alpha uv}(E)$,
$n^{+}_\alpha(E)$ the Fermi distribution
function of
$\alpha$-reservoir and $n^{-}_\alpha(E)=1-n^{+}_\alpha(E)$.

For later use, we introduce the dissipaton decomposition for the hybridizing bath \cite{Yan14054105}
in the energy-domain,
\begin{align}\label{dissp}
 \wti F^{\sigma}_{\alpha u} \equiv \bar\sigma    F^{\sigma}_{\alpha u}
 \equiv \!\int_{-\infty}^{\infty}\!\!\frac{{\mathrm d} E}{2\pi}
 f^{\sigma}_{\alpha u}(E).
\end{align}
The so-called dissipatons \{$f^{\sigma}_{\alpha u}(E)$ \} satisfy
\begin{align*}\label{all_notation}
\la f^{\sigma}_{\alpha u }(E,t)f^{\bar\sigma}_{\beta v }(E',0)\ra
   = - \delta_{\alpha\beta}\delta(E-E')
 e^{\sigma i E t}\Gamma^\sigma_{\alpha u v}(E).
\end{align*}
 It is easy to verify that the above decomposition preserves
the bath correlation function given by \Eq{ct}.

\subsection{TNL-ME and TL-EOM}
\label{thnmkme}

Let us outline the
TNL-ME and the equivalent energy-dispersed TL-EOM for weak system-reservoir coupling.
It is well-known that the primary central system
 is described by the reduced density operator, $\rho(t)\equiv{\rm tr}_{\B}[\rho_{\T }(t)]$,
i.e., the partial trace of the total density operator $\rho_{\T}$ over the bath
space. The corresponding
dynamics is determined by the TNL-ME,
 $\dot\rho(t) = -i[H_{\tS},\rho(t)]
  - \int_{t_0}^t\!\!{\mathrm d}\tau
   \Sigma(t-\tau)\rho(\tau)$.
 It can describe the non-Markovian dynamics for 
the self-energy $\Sigma(t-\tau)$ containing the memory effect.
Assuming weak system-bath coupling and performing Born but without
Markovian approximation,
the self-energy for the expansion up to second-order
of the tunneling Hamiltonian is expressed as
$\Sigma(t-\tau)=\big\la{\cal L}'(t) e^{-i{\cal L}_{\tS}(t-\tau)}{\cal L}'(\tau)
 \big\ra_{\B}$
in the $H_{\B}$-interaction picture.
The resulted TNL-ME is explicitly given by:
\be\label{TNL-ME}
\dot\rho(t)= -i{\cal L}_{\tS}\rho(t) -i\sum_{\alpha u\sigma}
  \big[a^{\bar\sigma}_u, \varrho^{\sigma}_{\alpha u}(t)\big],
 \ee
with ${\cal L}_{\tS} \hat O=[H_{\tS},\hat O]$ and
\be\label{TNL-ME1}
 \varrho^{\sigma}_{\alpha u}(t)
= -i\int_{t_0}^t\!\!{\mathrm d}\tau\, e^{-i{\cal L}_{\tS} (t-\tau)}
 {\cal C}^{(\sigma)}_{\alpha u}(t-\tau) \rho(\tau),
\ee
where
\be\label{calCt}
 {\cal C}^{(\sigma)}_{\alpha u}(t) \hat O
 \equiv \sum_{v}   \big[C^{(\sigma)}_{\alpha uv}(t)a^\sigma_{v}\hat O
 - C^{(\bar\sigma)\ast}_{\alpha uv}(t)\hat O a^{\sigma}_v\big].
 \ee
This depends on the bath correlation function, \Eq{ct}.
In \Eq{TNL-ME}, the first term describes the intrinsic coherent dynamics
and the second term
depicts the non-Markovian dissipative effect of the coupled reservoirs.

 Let $\rho(t)\equiv \Pi(t,t_0)\rho(t_0)$ be
the formal solution to \Eq{TNL-ME}.
Note that $\Pi(t,t_0)\neq\Pi(t,t_1)\Pi(t_1,t_0)$.
In other words, the conventional quantum regression theorem
is not directly applicable for
the calculation of the correlation functions.
 Alternatively, with the introduction of
${\bm\rho}(t)
\!\equiv\!\left[\rho(t),\rho^{\pm}_{\alpha u}(E,t) \right]^T$,
 the TNL-ME (\ref{TNL-ME}), with \Eq{TNL-ME1},
 can be converted to TL-EOM \cite{Jin16083038}
  \bsube\label{TL-EOM}
 \begin{align}
 \dot\rho(t)
 &=\!-i{\cal L}_{\tS}\rho(t)-\!i\!\sum_{\alpha u\sigma}\!\int\! \frac{{\rm d}E}{2\pi}
 \big[ a^{\bar\sigma}_u,\rho^{\sigma}_{\alpha u}(E,t)\big],
\label{rho0t}
\\
\dot\rho^{\sigma}_{\alpha u}(E,t)
&=\!-i({\cal L}_{\tS}\!-\!\sigma E)\rho^{\sigma}_{\alpha u}(E,t)
  -i{\cal C}^{(\sigma)}_{\alpha u}(E) \rho(t),
 \label{rho1t}
 \end{align}
\esube
where
 ${\cal C}^{(\sigma)}_{\alpha u}(E)=\int\! {\rm d}t\, e^{-\sigma iEt} {\cal C}^{(\sigma)}_{\alpha u}(t)$;
  cf.\,\Eq{calCt},
 \be\label{calCw}
 {\cal C}^{(\sigma)}_{\alpha u}(E) \hat O\equiv\sum_v \left[\Gamma^{(\sigma)}_{\alpha u v}(E)
 a^{\sigma}_v \hat O-\hat O \Gamma^{(\bar\sigma)\ast}_{\alpha u v}(E)a^{\sigma}_v\right].
 \ee
As implied in \Eq{TNL-ME1}, we have
  \begin{align}\label{varrho-phi}
 \varrho^{\sigma}_{\alpha u}(t) = \int\frac{{\rm d}E}{2\pi}\rho^{\sigma}_{\alpha u}(E,t).
 \end{align}
%

 Equation (\ref{TL-EOM}) can be summarized as
$\dot {\bm\rho}(t)={\bf{\Lambda}}\bm{\rho}(t)$
which leads to the solution of ${\bm\rho}(t)=\bm\Pi(t,t_0)\bm{\rho}(t_0)$ with
$\bm\Pi(t,t_0)=e^{{\bm\Lambda}(t-t_0)}$.
The TL-EOM space propagator satisfies
the time translation invariance, i.e., $\bm\Pi(t,t_0)=
\bm\Pi(t,\tau)\bm\Pi(\tau,t_0)$.
In other words, the TL-EOM \Eq{TL-EOM} is a mathematical
isomorphism of the conventional ``Schr\"{o}dinger equation''
and applicable to any physically supported initial state $\rho_{\T}(t_0)$.
In particular, the total system-plus-bath composite density operator $\rho_{\T}(t)$ maps to
 ${\bm{\rho}}(t)$, including the nonequilibrium steady state mapping,
 $\rho^{\rm st}_{\T} \rightarrow  {\bm{\rho}}^{\rm st}$.
This protocol can be extended to system correlation functions and
current correlation functions.
This is the advantage of TL-EOM (\ref{TL-EOM}) over TNL-ME (\ref{TNL-ME}).
The details are as follows.

\section{Current and noise spectrum}
\label{thcurr}

\subsection{The current formula}
\label{thsubcurr}

First, we identify
$\rho^{\sigma}_{\alpha u}( E,t)$ in \Eq{TL-EOM}
being the current-related density operator.
By the definition, the lead-specified current operator is
$\hat I_{\alpha}=-{\rm d}\hat N_{\alpha}/{\rm d}t=-i[\hat N_{\alpha},H']$,
with $\hat N_{\alpha}\equiv\sum_k c^\dg_{\alpha k}c_{\alpha k}$
being the number operator.
The tunneling Hamiltonian $H'$ is given by \Eq{Hsb1} with \Eq{dissp}.
We immediately obtain
\begin{align}\label{currI_hat}
  \hat I_{\alpha}
&= -i \sum_{\sigma u}  \ti a^{\sigma}_{ u}
    {\wti F}^{\bar\sigma}_{\alpha u}
  =-i\sum_{\sigma u}\int\! \frac{{\mathrm d} E}{2\pi}  \ti a^{\sigma}_u
  f^{\bar\sigma}_{\alpha u}( E),
\end{align}
where $\ti a^{ \sigma}_{ u}\equiv \sigma a^{\sigma}_{ u}$.
The average current reads
\begin{align}
  I_{\alpha}(t)
  &\!=\!{\rm Tr}[\hat I_{\alpha}\rho_{\T}(t)]
  \!=\!-i\!\sum_{\sigma u}\! \int\! \frac{{\mathrm d} E}{2\pi}{\rm tr}_{\rm s}
   [ \ti a^{\sigma}_{ u}\rho^{\bar\sigma}_{\alpha u }( E,t)],
\label{curr-ddo}
\end{align}
where
\begin{align}
\rho^{\sigma}_{\alpha u}( E,t)
 &\equiv
  {\rm tr}_{\B}\big[f^{\sigma}_{\alpha u}( E)\rho_{\T}(t)\big].
\label{phi1}
\end{align}
On the other hand,
performing the bath subspace trace (${\rm tr}_{\B}$) over
$\dot{\rho}_{\T}(t)=-i[H_{\tS}+H_{\B}+H',\rho_{\T}(t)]$,
 we obtain immediately \Eq{rho0t}, where $\rho^{\sigma}_{\alpha u}( E,t)$
 is the right given by \Eq{phi1}.
 In other words,
 TL-EOM (\ref{TL-EOM}) provides not only the real-time dynamics, but also transient current,
\Eq{curr-ddo}, with \Eqs{varrho-phi} and (\ref{TNL-ME1}),
\be\label{curr-exp}
  I_{\alpha}(t)
=-\!\sum_{\sigma u} \!\!\int_{t_0}^t\!\! {\mathrm d}\tau\, {\rm tr}_{\rm s}
   [ \ti a^{\sigma}_{ u}e^{-i{\cal L}_{\tS} (t-\tau)}{\cal C}^{(\bar\sigma)}_{\alpha u}(t-\tau) \rho(\tau) ].
\ee
Here, we set $\rho^{\pm}( E,t_0\!\rightarrow\!-\infty)=0$ for the initially
 decoupled system and reservoir.

\subsection{Current correlation function}
\label{thsubcurrcf}

Turn to the lead-specified steady-state current correlation function,
\begin{align}\label{CorrI}
 \la \hat I_{\alpha}(t)\hat I_{\alpha'}\!(0)\ra
 &={\rm Tr}\big[\hat I_{\alpha} \rho_{\T}(t; {\alpha'})\big],
\end{align}
with
\be
\rho_{\T}(t; {\alpha'})=  e^{-i{\cal L}_{\T}t} (\hat I_{\alpha'}\rho^{\rm st}_{\T})
\equiv e^{-i{\cal L}_{\T}t} \rho_{\T}(0; {\alpha'}).
\ee
Its TL-EOM correspondence reads
\be
\bm\rho (t; {\alpha'})=  e^{{\bm\Lambda}t}(\hat I_{\alpha'} \bm\rho^{\rm st} )
\equiv e^{{\bm\Lambda}t}{\bm\rho}(0; {\alpha'}).
\ee
Here,
$\bm{\rho}(t;\alpha')
\!\equiv\!\left[\rho(t;\alpha'),\rho^{\pm}_{\alpha u}(E,t;\alpha')\right]^T$,
with the propagator being defined in \Eq{TL-EOM}
and the initial values via \Eq{curr-ddo} being
\bsube\label{vecI02}
\begin{align}
\label{rho0alpha}
&\rho(0;\alpha')
 \equiv {\rm tr}_{\B} \big(\hat I_{\alpha'}\rho^{\rm st}_{\T}\big)
=-i\!\sum_{\sigma u}\! \int\! \frac{{\rm d} E}{2\pi}
     \ti a^{\bar\sigma}_{ u}\bar\rho^{\sigma}_{\alpha' u }(E) ,
\\
&\rho^{\sigma}_{\alpha u}(E,0;\alpha')
\equiv{\rm tr}_{\B}\big[f^{\sigma}_{\alpha u}( E)
\hat I_{\alpha'}\rho^{\rm st}_{\T}\big]
\nl&\hspace{5.5 em}
= -i\delta_{\alpha\alpha'}\!\sum_v \Gamma^{\sigma}_{\alpha u v}(E)
  \ti a^{\sigma}_{ v}\bar\rho,
\label{phiI0}
\end{align}
\esube
where $\bar\rho\equiv \rho^{\rm st}$ and
$\bar\rho^{\sigma}_{\alpha' u }(E)
\equiv [\rho^{\sigma}_{\alpha' u }(E)]^{\rm st}$.
We can then evaluate \Eq{CorrI} as
\begin{align}\label{CorrI1}
 \la \hat I_{\alpha}(t)\hat I_{\alpha'}\!(0)\ra
&=-i\sum_{\sigma u}\!\int\! \frac{{\mathrm d} E}{2\pi}{\rm tr}_{\rm s}
   [ \ti a^{\sigma}_{ u}\rho^{\bar\sigma}_{\alpha u }( E,t;\alpha')].
\end{align}

\subsection{Quantum noise spectrum}
\label{thsubnoise}

The lead--specified shot noise spectrum is given by
\be\label{Sw0}
 S_{\alpha\alpha'}(\omega)=\int_{-\infty}^{\infty}\!\!{\rm d}t\,
  e^{i\omega t}  \La \delta{\hat I}_\alpha(t)
  \delta{\hat I}_{\alpha'}(0)\Ra,
\ee
with $\delta{\hat I}_\alpha(t)\equiv{\hat I}_\alpha(t)-I^{\rm st}_{\alpha}$;
i.e.,
\be\label{corr-curr}
  \La \delta{\hat I}_\alpha(t)\delta{\hat I}_{\alpha'}(0)\Ra
=\La {\hat I}_\alpha(t){\hat I}_{\alpha'}(0)\Ra
  -I^{\rm st}_{\alpha}I^{\rm st}_{\alpha'}.
\ee
The steady--state current,
$I^{\rm st}_{\alpha}\equiv {\rm Tr}(\hat I_{\alpha}\bar\rho_{\T})$,
satisfies
$I^{\rm st}_{\rm L}=-I^{\rm st}_{\rm R}$.
 To proceed, we apply the initial values, \Eq{vecI02},
and express \Eq{CorrI1} in terms of
\begin{align}
 &\la \hat I_{\alpha}(t)\hat I_{\alpha'}\!(0)\ra
=\delta_{\alpha\alpha'}\!\sum_{\sigma u v}\!
 {\rm tr}_{\rm s}[a^{\sigma}_{ u}e^{-i{\cal L}_{\tS} t}
  C^{(\bar\sigma)}_{\alpha uv}(t)a^{\bar\sigma}_{v} \bar\rho]
\nl&\quad
  -\sum_{\sigma u} \!\int_{t_0}^t\! {\mathrm d}\tau\,
 {\rm tr}_{\rm s}[\ti a^{\sigma}_{ u} e^{-i{\cal L}_{\tS} (t-\tau)}
   {\cal C}^{(\bar\sigma)}_{\alpha u}(t-\tau) \rho(\tau;\alpha') ].
\label{curr-curr}
\end{align}
As detailed in Appendix,
the first term describes the contribution from \Eq{phiI0},
the second term involves $\rho(\tau;\alpha')$,
with the initial value of \Eq{rho0alpha}.

 To resolve $\rho(\tau;\alpha')$,
one can exploit either TNL-ME (\ref{TNL-ME})
or TL-EOM (\ref{TL-EOM}).
The related resolvent reads
\be
 \Pi(\omega)=[i({\cal L}_{\tS}-\omega)+\Sigma(\omega)]^{-1},
\ee
with
 $\Sigma(\omega)=\sum_{\alpha}
\big[{\cal J}^{<}_{\alpha}(\omega)-{\cal J}^{>}_{\alpha}(\omega)\big]$,
\bsube\label{caljomega}
 \begin{align}
{\cal J}^{>}_{\alpha}(\omega)\hat O&\equiv\!-\!\sum_{\sigma u}\ti a^{\bar\sigma}_{ u}
\big[{\cal C}^{(\sigma)}_{\alpha u}(\omega-{\cal L}_{\tS})\hat O\big],
\\
{\cal J}^{<}_{\alpha}(\omega)\hat O&\equiv\!-\!\sum_{\sigma u}
\big[{\cal C}^{( \sigma)}_{\alpha u}(\omega-{\cal L}_{\tS})\hat O\big]\ti a^{\bar\sigma}_{ u},
\end{align}
\esube
where [cf.\,\Eq{calCt}]
\begin{align}\label{calCw}
 {\cal C}^{(\sigma)}_{\alpha u}(\omega)  \hat O
 \!=\!\! \sum_{v}
 \! \big[C^{(\sigma)}_{\alpha uv}(\omega)(a^\sigma_{v}\!\hat O)
 \!- C^{(\bar\sigma)\ast}_{\alpha uv}( -\omega)(\hat O a^{\sigma}_v)\big],
 \end{align}
and $C^{(\sigma)}_{\alpha u v}( \omega)\equiv
\int_{0}^{\infty}\!dt\,
e^{i\omega t}C^{(\sigma)}_{\alpha u v}(t)$.
Denote further
\bsube\label{calwomega}
 \begin{align}
{\cal W}^{>}_{\alpha}(\omega)\hat O&\equiv\sum_{ \sigma uv}
\big[ \ti a^{\bar\sigma}_{u},C^{(\sigma)}_{\alpha uv }(\omega-{\cal L}_{\tS})
   (a^\sigma_{v}\hat O) \big],
\\
{\cal W}^{<}_{\alpha}(\omega)\hat O&\equiv
 \sum_{ \sigma uv}  \big[ \ti a^{\bar\sigma}_{u},C^{(\bar\sigma)\ast}_{\alpha uv }({\cal L}_{\tS}-\omega)
   (\hat O a^{\sigma}_{v})\big].
\end{align}
\esube
Note that
\bsube\label{caljw}
\begin{align}
 \big[{\cal W}^{<}_{\alpha}(\omega)\hat O\big]^\dg
&={\cal W}^{>}_{\alpha}(-\omega)\hat O^\dg,
\\
 \big[{\cal J}^{<}_{\alpha}(\omega)\hat O\big]^\dg
&={\cal J}^{>}_{\alpha}(-\omega)\hat O^\dg.
\end{align}
\esube
Moreover,
we have $I^{\rm st}_{\alpha}
={\rm tr}_{\tS}\big[{\cal J}^{>}_{\alpha}(0)\bar{\rho}\big]
={\rm tr}_{\tS}\big[{\cal J}^{<}_{\alpha}(0)\bar{\rho}\big]$,
for the steady current,
as inferred from \Eq{curr-exp}.

Finally, we obtain \Eq{Sw0},
with \Eqs{corr-curr} and (\ref{curr-curr}),
the expression
(see Appendix for the derivations),
\begin{align}\label{Sw}
S_{\alpha\alpha'}(\omega)
 &= {\rm tr}_{\rm s}\Big\{{\cal J}^{>}_{\alpha}(\omega)
 \Pi(\omega)\big[{\cal J}^{>}_{\alpha'}(0)
   + {\cal W}^{>}_{\alpha'}(\omega)\big]\bar\rho
\nl&\qquad
+{\cal J}^{<}_{\alpha'}(-\omega)\Pi(-\omega)\big[{\cal J}^{<}_{\alpha}(0)  +
  {\cal W}^{<}_{\alpha}(-\omega) \big]\bar\rho\Big\}
\nl&\quad
 +2\delta_{\alpha'\alpha}{\rm Re}\! \sum_{\sigma u v}
   {\rm tr}_{\rm s}\big[ a^{\bar\sigma}_u  C^{(\sigma)}_{\alpha uv }(\omega-{\cal L}_{\tS})
        a^{\sigma}_{v}{\bar\rho}
          \big].
 \end{align}
This is the key result of this paper,
with $\omega>0$ and $<0$
corresponding to energy
absorption and emission processes,
respectively \cite{Eng04136602,Agu001986,Jin15234108,Nak19134403}.

\subsection{Discussions and remarks}
\label{thRemarks}



In mesoscopic quantum transport,
the charge conservation is about
$-\dot{Q}(t)=I_{\rm L}(t)+I_{\rm R}(t)\equiv I_{\rm dis}(t)$,
with the displacement current arising from the change of the charge $Q(t)$
in the central system. The corresponding
fluctuation spectrum,
$S_{\rm c}(\omega)=\int_{-\infty}^{\infty} \!dt\,
  e^{i\omega t} \La \delta{\dot{Q}(t)} \delta{\dot{Q}(0)}\Ra$,
 can then be evaluated via \cite{Jin11053704}
\begin{align}\label{Scw}
S_{\rm c}(\omega)&=S_{\rm LL}(\omega)+S_{\rm RR}(\omega)+2{\rm Re}[S_{\rm LR}(\omega)].
\end{align}
 For auto-correlation noise spectrum,
\Eq{Sw} with $\alpha'=\alpha$,
we have [\cf\Eq{caljw}]
\begin{align}\label{Sw-auto}
S_{\alpha\alpha}(\omega)
 &=2\,{\rm Re}\,{\rm tr}_{\rm s}\big\{{\cal J}^{>}_{\alpha}(\omega) \Pi(\omega)\big[{\cal J}^{>}_{\alpha}(0)
   + {\cal W}^{>}_{\alpha}(\omega)\big]\bar\rho\big\}
\nl&\quad
 +2\,{\rm Re}\sum_{\sigma u v}
   {\rm tr}_{\rm s}\big[ a^{\bar\sigma}_u  C^{(\sigma)}_{\alpha uv }(\omega-{\cal L}_{\tS})
        a^{\sigma}_{v}{\bar\rho}\big].
  \end{align}
Alternatively, $S_{\rm c}(\omega)$ can also be calculated via $S_{\rm c}(\omega)=e^2\omega^2 S_{\rm N}(\omega)$,
where $S_{\rm N}(\omega)\equiv {\cal F}[\delta \hat N(t)\delta \hat N(0)]$,
with $\hat N =\sum_u a^\dg_u a_u$. The spectrum of the charge fluctuation $S_{\rm N}(\omega)$
can be evaluated straightforwardly by the established formula for the non-Markovian
 correlation function of the system operators in our previous work \cite{Jin16083038}.

The total current in experiments reads
 $I(t) =a I_{\rm L}(t)- bI_{\rm R} (t)$,
 with the junction capacitance parameters ($a,b\geq0$) satisfying $a+b=1$
  \cite{Bla001,Wan99398,Mar10123009}.
In wide-band limit, $a=\frac{\Gamma_{\rm R}}{\Gamma_{\rm L}+\Gamma_{\rm R}}$ 
 and $b=\frac{\Gamma_{\rm L}}{\Gamma_{\rm L}+\Gamma_{\rm R}}$ \cite{Wan99398}.
The total current noise spectrum
can be calculated via either
 \be\label{Swtotal}
 S(\omega) = a^2S_\text{LL}(\omega)+b^2S_\text{RR}(\omega)
-2ab\,{\rm Re}[S_\text{LR}(\omega)],
\ee
or
\be\label{Swtotal2}
 S(\omega) = aS_\text{LL}(\omega)+bS_\text{RR}(\omega)
-ab\,S_{\rm c}(\omega).
\ee
%

As known, the present method is a second-order theory and applicable for weak
system-reservoir coupling, i.e., $\Gamma\lesssim k_{\rm B}T$. This describes the
electron sequential tunneling (ST) processes.
The resulted noise formula expressed by \Eq{Sw}
in principle is similar to that obtained in Ref.\,\onlinecite{Eng04136602}.
The most advantage of [\Eq{Sw}]
is that the involved supertoperators have well-defined in \Eq{caljomega}
and \Eq{calwomega}.
One only needs the matrix operations where
 we should
 transform the Liouville operator ${\cal L}_{\tS}$ into energy difference
 in the eigenstate basis $\{|n\ra\}$ ($H_{\tS}|n\ra=\varepsilon_n|n\ra$), e.g.,
 $\la n|f({\cal L}_{\tS})\hat Q|m\ra=f(\varepsilon_n-\varepsilon_m)Q_{nm}$.
%

In \Eq{Sw},
the memory effect enters through
the frequency--dependence in the last term and also in ${\cal J}^{\lgter}_{\alpha}(\omega)$
and ${\cal W}^{\lgter}_{\alpha}(\omega)$. In the Markovian limit, \Eq{Sw} reduces to
\begin{align}\label{Swmk}
S^{\rm Mar}_{\alpha\alpha'}(\omega)
 &= {\rm tr}_{\rm s}\Big\{{\cal J}^{>}_{\alpha}(0)
 \Pi_0(\omega)\big[{\cal J}^{>}_{\alpha'}(0)
   + {\cal W}^{>}_{\alpha'}(0)\big]\bar\rho
\nl&\qquad
+{\cal J}^{<}_{\alpha'}(0)\Pi_0(-\omega)\big[{\cal J}^{<}_{\alpha}(0)  +
  {\cal W}^{<}_{\alpha}(0) \big]\bar\rho\Big\}
\nl&\quad
 +2\delta_{\alpha'\alpha}{\rm Re}\! \sum_{\sigma u v}
   {\rm tr}_{\rm s}\big[ a^{\bar\sigma}_u  C^{(\sigma)}_{\alpha uv }(-{\cal L}_{\tS})
        a^{\sigma}_{v}{\bar\rho}
          \big],
 \end{align}
where $\Pi_0(\omega)=[i({\cal L}_{\tS}-\omega)+\Sigma(0)]^{-1}$
with $\Sigma(0)=\sum_{\alpha}
\big[{\cal J}^{<}_{\alpha}(0)-{\cal J}^{>}_{\alpha}(0)\big]$.
The involved superoperators were defined in \Eq{caljomega}
 and \Eq{calwomega}.
The widely studied Markovain problems \cite{Xu02023807,Li04085315,Li05205304,Li05066803,Luo07085325,Mar10123009}
had also considered the Redfield approximation with
the neglect of the bath dispersion $\Lambda^{(\pm)}_{\alpha uv}(\omega)$
in \Eq{appcw} (the imaginary part of $C^{(\pm)}_{\alpha uv}(\omega)$).
 One then can easily check that
${\rm Re}[S^{\rm Mar}_{\alpha\alpha'}(\omega)]={\rm Re}[S^{\rm Mar}_{\alpha'\alpha}(-\omega)]$ with $\alpha\neq\alpha'$
and $S^{\rm Mar}_{\alpha\alpha}(\omega)=S^{\rm Mar}_{\alpha\alpha}(-\omega)$ based on \Eq{Swmk}.
In other words, Markovian transport corresponds to
the symmetrized spectrum.

\section{Numerical demonstrations}
\label{thnum}

To verify the validity of the established method, 
we will apply it to demonstrate the quantum noise spectrum of the transport
current through interacting double quantum dots (DQDs).
All the numerical results will be further compared with exact results based on
DEOM theory.

The total composite Hamiltonian of the DQDs
 contacted by the two electrodes is described by \Eq{Htot0}.
The Hamiltonian for the DQDs in series is specified by,
\be\label{Hs-cqd}
 H_{\tS}= \varepsilon_{l}a^\dg_la_l + \varepsilon_{r}a^\dg_ra_r
   +U \hat n_l \hat n_r+\Omega\big(a^\dg_{l} a_{r}+a^\dg_{r} a_{l}\big).
\ee
where $U$ is the inter-dot Coulomb interaction, $\Omega$ describes the
inter-dot electron coherent transition,
 and $\hat n_u=a^\dg_u a_u$.
The involved states of the double dot are $|0\ra$ for the empty double dot,
$|l\ra$ for the left dot occupied, $|r\ra$
for the right dot occupied, and $|2\ra\equiv|lr\ra$ for the two dots occupied.
Under the assumption of
the infinite intra Coulomb interaction and large Zeeman
split in each dot,
we consider at most one electron in each dot.
In this space, we have $a_{l}=|0\ra\la l|+|r\ra\la2|$
and $a_{r}=|0\ra\la r|-|l\ra\la 2|$.
Apparently, the single-electron occupied states
of $|l\ra$ and $|r\ra$ are not the eigenstates of the
system Hamiltonian $HS_{\tS}$.
It has the intrinsic coherent Rabi oscillation
demonstrated by the coherent coupling strength $\Omega$.
The corresponding Rabi frequency denoted by $\Delta$ is
the energy difference between the two eigenstates ($\varepsilon_{\pm}$),
e.g., $\Delta=\varepsilon_{+}-\varepsilon_{-}=2\Omega$ for
the degenerate DQDs ($\varepsilon_{l}=\varepsilon_{r}=\varepsilon_{0}$) considered here.
The characteristic of the Rabi coherence has been well studied in the symmetrized noise spectrum
\cite{Luo07085325,Agu04206601,Mar11125426,Shi16095002}.

Now we apply the present TL-EOM approach
to calculate the quantum noise spectrums
of the transport current through DQDs.
As we mentioned above, the TL-EOM method is suitable for weak system-reservoir
coupling which can appropriately
describe the electron ST processes.
We thus consider the ST regime
where the energy levels in DQDs are within the bias
window ($\mu_{\rm L}>\varepsilon_{0}>\mu_{\rm R}$).
Without loss of generality,
we set antisymmetry bias voltage with $\mu_{\rm L}=-\mu_{\rm R}=eV/2$
and the energy level with $\varepsilon_{0}=0$.
The wide band width is considered
with setting $W_{\alpha}= 300\Gamma$
in \Eq{jw}.
 We adopt the total coupling strength of $\Gamma=\Gamma_{\rm L}+\Gamma_{\rm R}$ as the unit of
 the energy and focus on the symmetrical coupling strength
  $\Gamma_{\rm L}=\Gamma_{\rm R}=0.5\Gamma$ (a=b=1/2)
in this work.
 Furthermore, we test the upper limit of the system-reservoir coupling
which is comparable to the order of the temperature ($\Gamma\approx k_{\rm B}T$), with setting
$k_{\rm B}T=0.5\Gamma$ here.
Details for the other parameters are given in the figure captions.

\begin{figure}
\includegraphics[width=1.0\columnwidth]{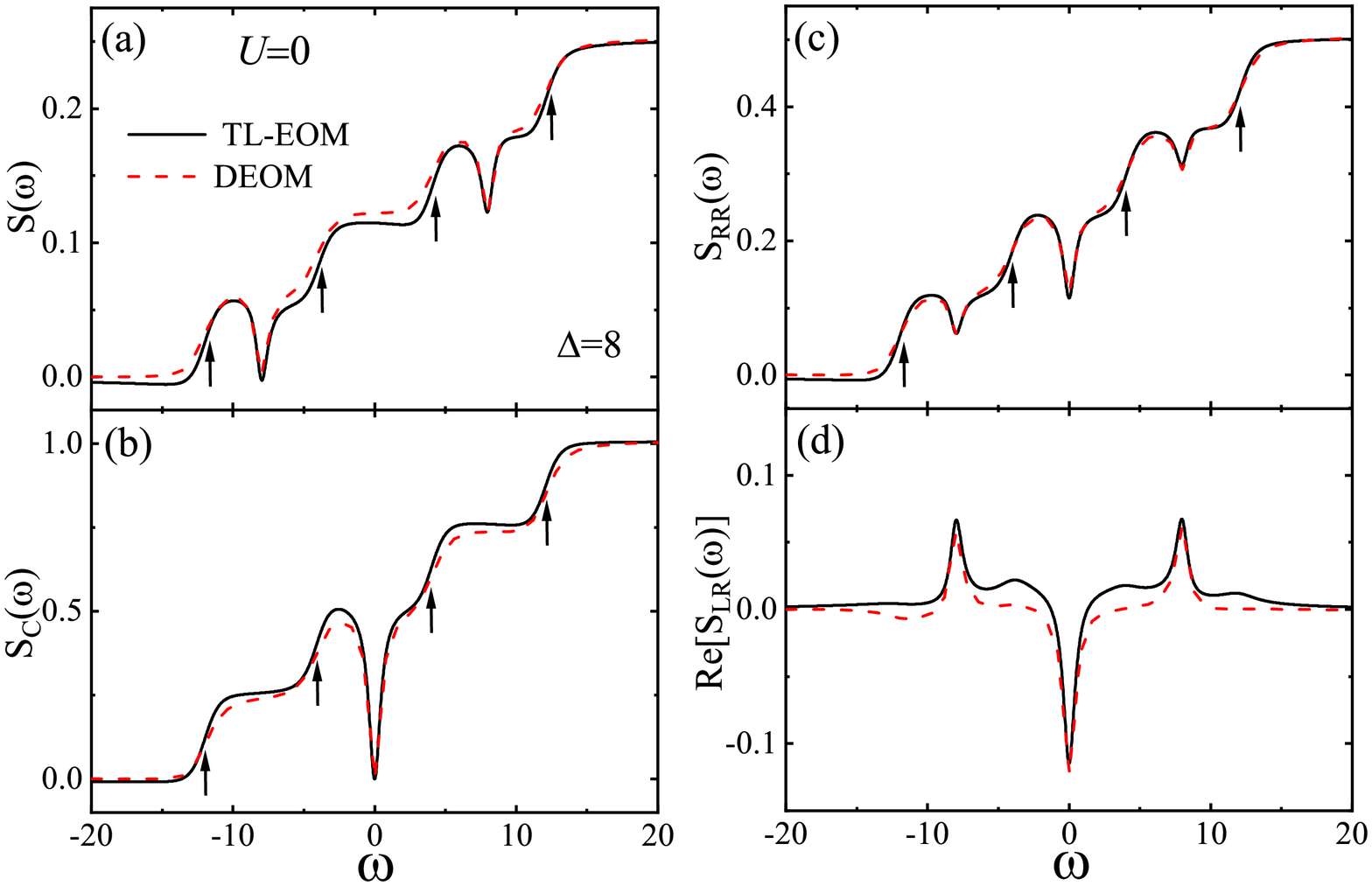}
\caption{(Color online)
The total and the lead-specified current noise spectra with noninteracting effect ($U=0$)
based on TL-EOM method (black-solid line) and exact DEOM theory (red-dash line).
(a) The total current noise spectrum, $S(\omega)$.
(b) The central current fluctuation spectrum, $S_{\rm c}(\omega)$.
(c) The auto-correlation noise spectrum of $R$-lead, $S_{\rm RR}(\omega)$.
(d) The cross-correlation noise spectrum, ${\rm Re}[S_{\rm LR}(\omega)]$.
  The other parameters (in unit of $\Gamma$) are $\Omega=4$
  and $eV=16$.
}
\label{fig1}
\end{figure}

\begin{figure}
\includegraphics[width=1.0\columnwidth,angle=0]{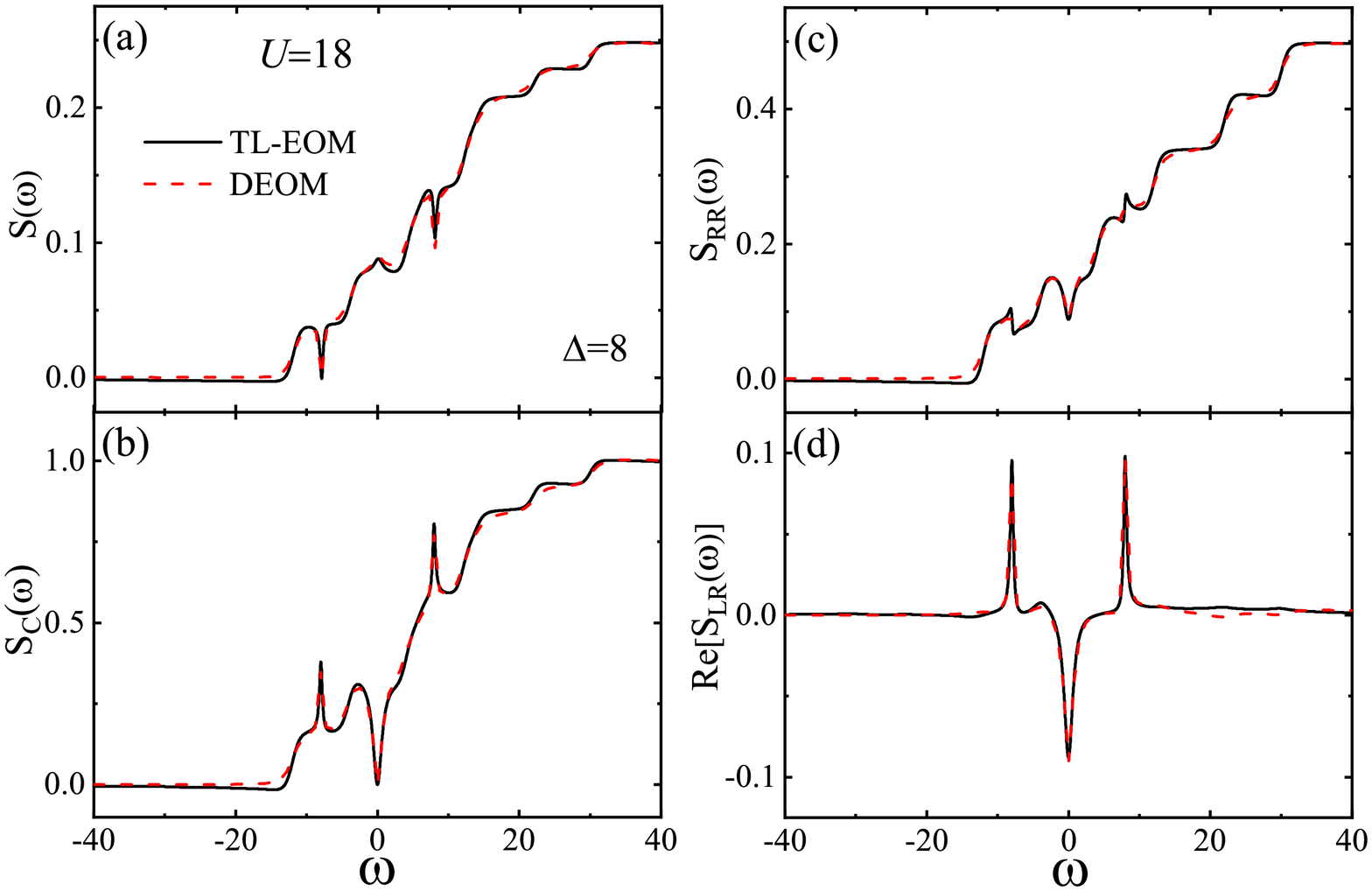}
\caption{(Color online)
The total and the lead-specified current noise spectra with
with strong inter-dot Coulomb interaction ($U=18\Gamma$)
based on TL-EOM method (black-solid line) and exact DEOM theory (red-dash line).
(a) The total current noise spectrum, $S(\omega)$.
(b) The central current fluctuation spectrum, $S_{\rm c}(\omega)$.
(c) The auto-correlation noise spectrum of $R$-lead, $S_{\rm RR}(\omega)$.
(d) The cross-correlation noise spectrum, ${\rm Re}[S_{\rm LR}(\omega)]$.
 The other parameters are the same as in \Fig{fig1}. }
  \label{fig2}
\end{figure}

\begin{figure}
\includegraphics[width=1.0\columnwidth,angle=0]{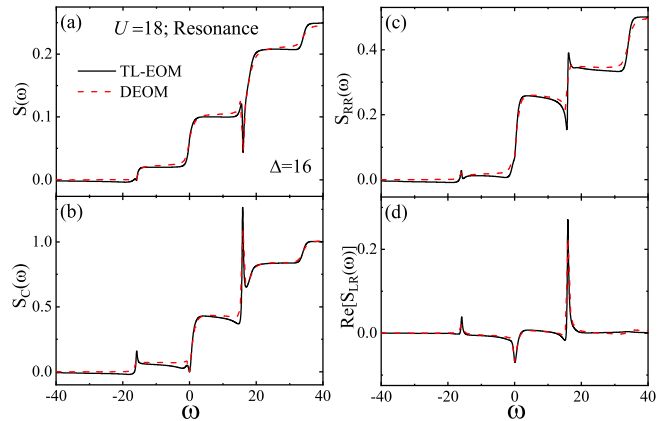}
\caption{(Color online)
The total and the lead-specified current noise spectra at resonance regime
($\varepsilon_{\pm}=\pm\Omega=\pm8\Gamma=\pm eV/2$)
with strong inter-dot Coulomb interaction ($U=18\Gamma$),
based on TL-EOM method (black-solid line) and exact DEOM theory (red-dash line).
(a) The total current noise spectrum, $S(\omega)$.
(b) The central current fluctuation spectrum, $S_{\rm c}(\omega)$.
(c) The auto-correlation noise spectrum of $R$-lead, $S_{\rm RR}(\omega)$.
(d) The cross-correlation noise spectrum, ${\rm Re}[S_{\rm LR}(\omega)]$.
 The other parameters are the same as in \Fig{fig1}. }
  \label{fig3}
\end{figure}

The numerical results
of the total and the lead-specified current noise spectra are
displayed in Figs.\,\ref{fig1}, \ref{fig2} and \ref{fig3}.
They correspond to noninteracting ($U=0$), strong inter-dot interacting ($U=18\Gamma$),
and the resonance regime ($U=18\Gamma$ and $\varepsilon_{+/-}=\mu_{\rm L/R}$), respectively.
Furthermore, the evaluations are based on
the present TL-EOM method (black solid-line) and the exact DEOM theory (red dash-line).
Evidently, the TL-EOM method reproduces well, at least qualitatively,
all the basic features of the quantum noise spectra in the entire frequency range.
The detail demonstrations see below.

Figures \ref{fig1} depicts the noise spectra
in the absence of the inter-dot Coulomb interaction ($U=0$).
The characteristics are as follow:
(\emph{i}) The well--known quasi-steps around
the energy resonances,
$\omega=\pm\omega_{\alpha \pm}\equiv\pm|\varepsilon_{\pm}-\mu_\alpha|$,
emerge in the total noise spectrum $S(\w)$, the displacement $S_{c}(\w)$
and the diagonal component, exemplified with $S_{\rm RR}(\w)$;
see the arrows in \Fig{fig1}(a)--(c).
The aforementioned feature
arises from the non-Markovian dynamics
of the electrons in $\alpha$--electrode
tunneling into and out of the DQDs,
accompanied by
energy absorption ($\omega>0$) and emission ($\omega<0$), respectively.
(\emph{ii}) In addition,
the Rabi resonance at $\omega=\pm\Delta\equiv \pm (\varepsilon_{+}-\varepsilon_{-})$ appears
in
$S(\omega)$ [\Fig{fig1}(a)] and $S_{\rm RR}(\omega)$ [\Fig{fig1}(c)] as dips,
whereas in
${\rm Re}[S_{\rm LR}(\omega)]$ [\Fig{fig1}(d)] as peaks.
On the other hand, in $S_{\rm c}(\omega)$,
the former aforementioned dips and peaks are accidently canceled out [see \Fig{fig1}(b)],
in the absence of Coulomb interaction ($U=0$).
%

Figure \ref{fig2} depicts the noise spectra in the presence of
  strong inter-dot Coulomb interaction ($U=18\Gamma$).
(\emph{iii}) In contrast to \Fig{fig1}(b),
now the displacement current noise spectrum $S_{\rm c}(\omega)$
 displays at $\omega=\pm\Delta$ the Rabi coherence [see \Fig{fig2}(b)].
While the Rabi peaks are enhanced in ${\rm Re}[S_{\rm LR}(\omega)]$
   [see \Fig{fig2}(d)],
  the original Rabi dips in \Fig{fig1}(c) become peak-dip profile
in $S_{\rm RR}(\omega)$ [see \Fig{fig2}(c)].
(\emph{iv}) Moreover, the Coulomb-assisted transport channels ($\varepsilon_{\pm}+U$)
produces new non-Markovian quasi-steps around $\omega=\pm
\omega_{\alpha {\rm u}\pm}\equiv\pm|\varepsilon_{\pm}+U-\mu_\alpha|$
in the total, displacement, and the auto-correlation current
noise spectra, as shown in \Fig{fig2}(a)--(c).
%

In \Fig{fig3}, we highlight the characteristics of the noise spectra
in the resonance regime ($\varepsilon_{+/-}=\mu_{\rm L/R}$)
by increasing the coherent coupling strength $\Omega$.
(\emph{v})
Compared with \Fig{fig2}, the Rabi signal
in the absorption noise spectrum 
at $\omega=\Delta$ is remarkably enhanced, while
the signal in the emission one at $\omega=-\Delta$
is negligibly small. The observation here had been explored
in the isolation
of competing mechanisms, such as the Kondo resonance
emission noise spectrum \cite{Bas12046802, Del18041412,Mao21014104}.

The above absorptive versus emissive feature can
  be understood in terms of steady occupation from the following two aspects:
(1) Away from the energy resonance ($\mu_{\rm L}>\varepsilon_{\pm}>\mu_{\rm R}$),
 the probabilities of single-electron occupied states are nearly the same,
$\bar\rho_{++}\cong\bar\rho_{--}$.
The resulting energy absorption and emission are equivalent in the noise spectrum.
%
(2) In the energy resonance ($\varepsilon_{+/-}=\mu_{\rm L/R}$) region,
the stationary state is very different.
The lower energy state occupation on $|-\ra$ is the majority, e.g., $\bar\rho_{--}\gg\bar\rho_{++}$.
Thus, the Rabi feature in absorption is much stronger than
that in emission noise.

\section{Summary}
\label{thsum}

In summary, we have presented an efficient TL-EOM approach for the quantum noise
spectrum of the transport current through interacting mesoscopic systems.
%
The established method is
based on the transformation of the second-order non-Markovian master equation described by
TNL-EM into the energy-dispersed
 TL-EOM formalism by introducing the current-related density operator.
The resulted analytical formula of the current noise spectrum
can characterize the nonequilibrium transport including electron-electron Coulomb interaction and
the memory effect.

We have demonstrated the proposed method in transport through interacting-quantum-dots system,
and find good agreement with the exact results under broad range of parameters.
The numerical calculations are based on both
the present TL-EOM method and exact DEOM theory.
We find that all the basic features of the lead-specified noise spectra in the entire frequency range,
including energy-resonance and Coulomb-assisted non-Markovian
quasi-steps, and the intrinsic coherent Rabi signal, at least qualitatively,
are reconciled well with the accurate results.
As a perturbative theory, the present TL-EOM is applicable in the
  weak system-reservoir coupling ($\Gamma\lesssim k_{\rm B}T$) regime,
  dominated by sequential tunneling processes.
 Other parameters such as the bias voltage and Coulomb interaction,
are rather flexible.

\acknowledgments
We acknowledge helpful discussions with
  X. Q. Li.
  The support from the Ministry of Science and Technology of China (No. 2021YFA1200103)
  and the Natural Science Foundation of China
(Grant No. 11447006) is acknowledged.

\appendix*

\section{Derivation of the noise spectrum}
 \label{thappsw}

In this appendix, we first give the derivation for the current correlation
function \Eq{curr-curr} and then its Laplace transformation resulting in
the current noise spectrum formula \Eq{Sw}.
 The details are below.

In the current correlation function expressed by \Eq{CorrI1},
 one needs to get the solution of
$\rho^{\sigma}_{\alpha u }( E,t;\alpha')$.
Based on \Eq{rho1t} with the initial condition of \Eq{phiI0},
we have
\begin{align}\label{phi0}
 &\rho^{\sigma}_{\alpha u}(E,t;\alpha')
=-i\, \delta_{\alpha\alpha'}e^{-i({\cal L}_{\tS}-\sigma E)t}
\!\sum_v \Gamma^{\sigma}_{\alpha u v}(E)
  \ti a^{\sigma}_{ v}\bar\rho
  \nl&\quad\quad
    -i \int_{0}^t\!\!{\mathrm d}\tau
  e^{-i({\cal L}_{\tS}-\sigma E )(t-\tau)}{\cal C}^{\sigma}_{\alpha u}( E)
  \rho(\tau;\alpha').
 \end{align}
This straightforwardly recasts \Eq{CorrI1} to \Eq{curr-curr}.

Next, we are going to get the current noise spectrum formula \Eq{Sw},
according to the definition of \Eq{Sw0}.
Following the procedure of Refs.\,\cite{Yan14054105,Jin15234108,Yan16110306},
the quantum noise spectrum can be calculated via
\be\label{Sw_alp}
  S_{\alpha\alpha'}(\omega)={\cal I}_{\alpha \alpha'}(\omega)
    +{\cal I}^{\ast}_{\alpha' \alpha}(\omega),
\ee
with the introduction of
  \be\label{Cw_alp}
 {\cal I}_{\alpha\alpha'}(\omega)\equiv \int_0^{\infty} \!dt\,
  e^{i\omega t} \la  {\hat I}_\alpha(t) {\hat I}_{\alpha'}(0)\ra \,.
\ee
In terms of the steady current correlation function given by \Eq{curr-curr},
we can straightforwardly have,
\bsube\label{calIw}
\begin{align}\label{calIw1}
 {\cal I}_{\alpha\alpha'}(\omega)&=
 \delta_{\alpha'\alpha} \sum_{\sigma u v}
   {\rm tr}_{\rm s}\big[ a^{\bar\sigma}_u  C^{(\sigma)}_{\alpha uv }(\omega-{\cal L})
        a^{\sigma}_{v}{\bar\rho}
          \big]
 \nl&\quad
               +{\rm tr}_{\rm s}
 \big[{\cal J}^{>}_{\alpha}(\omega) \rho(\omega;\alpha')\big]
  ,
\\
\label{calIw2}
 {\cal I}^\ast_{\alpha'\alpha}(\omega)
 &=
 \delta_{\alpha'\alpha} \sum_{\sigma u v}
   {\rm tr}_{\rm s}\big[  C^{(\bar\sigma)\ast}_{\alpha uv }(\omega+{\cal L})
        \bar\rho a^{\sigma}_{v}a^{\bar\sigma}_u\big]
 \nl&\quad
   + {\rm tr}_{\rm s}
 \big[{\cal J}^{<}_{\alpha'}(-\omega)\rho^\dg(\omega;\alpha)\big],
\end{align}
\esube
where ${\cal J}^{\lgter}(\omega)$ is defined in \Eq{caljomega}, and
$C^{(\pm)}_{\alpha uv}(\omega)
=\int^\infty_0 dt\, e^{i\omega t}C^{(\pm)}_{\alpha uv}(t)$
is obtained as \cite{Jin14244111,Shi16095002,Jin16083038}
\begin{align}\label{appcw}
C^{(\pm)}_{\alpha uv}(\omega)
=\frac{1}{2}\big[
 \Gamma^{(\pm)}_{\alpha uv}(\mp \omega)
 +i\Lambda^{(\pm)}_{\alpha uv }(\mp \omega)\big].
\end{align}
The real [c.f.\Eq{cw-real}] and the imaginary parts are the so-called bath
interaction spectrum and {\it dispersion}, respectively \cite{Xu029196}.
They are related via the Kramers-Kronig relations: 
\begin{align}
\label{Lamb}
\Lambda^{(\pm)}_{\alpha uv}(\omega)
&={\cal P}\int^\infty_{-\infty}\frac{{\mathrm d}\omega'}{2\pi}
\frac{1}{\omega\pm\omega'}\Gamma^{(\pm)}_{\alpha uv}(\omega)
\nl&=\frac{\Gamma_{\alpha uv}(\omega)}{2\pi}
\Bigg\{{\rm Re}\left[\Psi\left(\frac{1}{2}
+i\frac{\beta(\omega-\mu_\alpha)}{2\pi}\right)\right]
\nla
-\Psi\left(\frac{1}{2}+\frac{\beta W_\alpha}{2\pi}\right)
\mp\pi\frac{\omega-\mu_\alpha}{W_\alpha}\Bigg\},
\end{align}
where ${\cal P}$ denotes the principle value of the integral,
and $\Psi(x)$ is the digamma function. In \Eq{Lamb}, the second identity is from
the consideration of the Lorentzian-type form
of the hybridization spectral density, i.e.,
\be\label{jw}
J_{\alpha uv}(\omega)=\frac{\Gamma_{\alpha uv} W^2_{\alpha}}{(\omega-\mu_\alpha)^2+W^2_{\alpha}},
\ee
with the coupling strength $\Gamma_{\alpha uv}$ and the bandwidth $W_{\alpha}$
of lead-$\alpha$.

In \Eq{calIw}, the key is to get the solutions of $\rho(\omega;\alpha')$
and $\rho^\dg(\omega;\alpha)$.
%
Setting $\dot\rho^{\sigma}_{\alpha u}(E,t)=0$ in
  \Eq{rho1t}, we first get
the stationary solution of
\be\label{phis}
\bar\rho^{\sigma}_{\alpha u}(E)=\frac{ \sum_v\big[\Gamma^{\sigma}_{\alpha uv}(E)
   a^\sigma_v\bar\rho
   -\Gamma^{{\bar\sigma}\ast}_{\alpha uv}(E)\bar\rho
   a^\sigma_v\big]}{\sigma E-{\cal L}_{\tS}+i0^+}.
\ee
Inserting it into \Eq{rho0alpha}, we have
\begin{align}\label{rho0alpha1}
 \rho(0;\alpha')
   &=\!-\!\sum_{\sigma u}\ti a^{\bar\sigma}_{ u}
{\cal C}^{(\sigma)}_{\alpha' u}(-{\cal L}_{\tS})\bar\rho
 ={\cal J}^{>}_{\alpha'}(0)\bar\rho.
 \end{align}
 Now $\rho(\omega;\alpha')=\int_{0}^{\infty}{\rm d} te^{i\omega t}
 \rho(t;\alpha')$ can then be obtained by
either TNL-ME (\ref{TNL-ME}) with the initial condition of \Eq{rho0alpha1}
 or by TL-EOM (\ref{TL-EOM})
with
 ${\bm\rho}(0;\alpha')$ described by \Eq{rho0alpha1}
and \Eq{phiI0}.
One then can easily get
  \begin{align}\label{rhow}
 \rho(\omega;\alpha')&=\Pi(\omega)\Big\{{\cal J}^{>}_{\alpha'}(0)\bar\rho +
  {\cal W}^{>}_{\alpha'}(\omega)\bar\rho\Big\}.
 \end{align}
 Further with the relation of \Eq{caljw}, we obtain
  \begin{align}
   \rho^\dg(\omega;\alpha)&=\Pi(-\omega)\Big\{{\cal J}^{<}_{\alpha}(0)\bar\rho +
  {\cal W}^{<}_{\alpha}(-\omega)\bar\rho\Big\}.
 \end{align}
 Finally, based on \Eq{Sw_alp} to \Eq{calIw}, the general lead-specified formula
 of quantum noise spectrum, \Eq{Sw}, is straightforwardly obtained.

\end{CJK*}

\end{document}